# Classical Tunneling

**Arthur Cohn and Mario Rabinowitz**

*Electric Power Research Institure, Palo Alto, CA 94062*
Inquiries to: Mario Rabinowitz
*Armor Research, 715 Lakemead Way, Redwood City, CA 94062*
Mario715@earthlink.net

**Abstract**
   A classical representation of an extended body over barriers of height greater than the energy of the incident body is shown to have many features in common with quantum tunneling as the center-of-mass literally goes through the barrier.  It is even classically possible to penetrate any finite barrier with a body of arbitrarily low energy if the body is sufficiently long.  A distribution of body lengths around the de Broglie wavelength leads to reasonable agreement with the quantum transmission coefficient.

## 1. INTRODUCTION

Quantum mechanics gives the best representation known of the atomic and subatomic world.  Classical mechanics has been assumed to be incapable of representing tunneling and other phenomena which have been presumed to be uniquely in the quantum domain.  This paper is presented in the spirit of Gryzinski (1965, 1972, 1973a,b), who showed that a semiclassical understanding of the atom, atomic collisions , and molecular forces becomes more reasonable as the classical model is refined.  It is our goal to show that quantum tunneling is also amenable to properly constructed classical analogs.  Consideration of an object having length (rather than being a point mass) will yield a classical explanation or analog to the concept of tunneling.  Further consideration of a distribution of lengths will yield an analog or counterpart to the quantum tunneling coefficient.

There have been a number of quasiclassical approaches to quantum mechanics (utilizing mechanisms such as background fluctuations) in which the Schrödinger equation and the Klein-Gordon equation have been derived (Bohm, 1952; Aron, 1965, 1966; Nelson, 1966; Lehr and Park, 1977; Park *et al.*, 1980). Once this has been achieved, tunneling is introduced in the same way as in quantum mechanics with no classical insight as to how a body can get through a barrier when its energy is less than that of the barrier height. Bohm (1952) has some of the particles going over the barrier due to "violent fluctuations in the quantum mechanical potential." Others have particles hopping over the barrier due to thermal fluctuations.

To our knowledge, no one has previously shown that there is a direct classical analog to quantum tunneling. Texts typically state: "This possibility of going through potential barriers – called the tunnel effect – makes it possible to understand in terms of quantum mechanics a number of atomic phenomena that are inexplicable classically" (Rojansky, 1964). This paper will show that an extended body can yield a classical representation of tunneling which has many of the quantum mechanical features.

Our paradigm is that of "high jumping" rather than tunneling for any force field. In the case of a gravitational field, the process is like that of a high jumper whose center of gravity does not have to be raised to the height of the bar in order for the jumper to clear the bar. The high jumper's body does not need to have an energy greater than or equal to that of the jumper's entire weight times the barrier height, since those parts of the body already over the barrier drop to a lower potential energy as the body crosses the barrier. In essence, our view is that tunneling may also be regarded as high jumping of an extended body which can clear a barrier even when its energy is less than the potential energy of the barrier, if it can communicate with and be aided by the interaction on the other side of the barrier. The siphon analog is also appropriate. For a one-dimensional barrier, the body need not be flexible. In higher dimensions the extended body should be flexible, and we call it a "rope" for convenience.

## 2. QUANTUM MECHANICAL TUNNELING

Let us first establish some properties of quantum mechanical tunneling with which to compare the classical high-jumping results. It is noteworthy that the quantum mechanical representation for a point particle is essentially a wave equation. In terms of the quantum wave-particle duality, tunneling deals only with the wave nature of an object, as the absence of localization precludes dealing with the object's particle nature. A wave has distributed energy and hence distributed equalivalent mass. In the conventional interpretation of quantum mechanics this possible inconsistency with the concept of a point particle is circumvented by interpreting the wave intensity as a probability distribution for finding the point particle.

The stationary solution of tunneling is obtained by solving the Schrödinger equation on the incident side of the barrier, inside the barrier, and on the transmitted side of the barrier:

$$\frac{-\hbar^2}{2m}\nabla^2\Psi + (V - E)\Psi = 0 \tag{1}$$

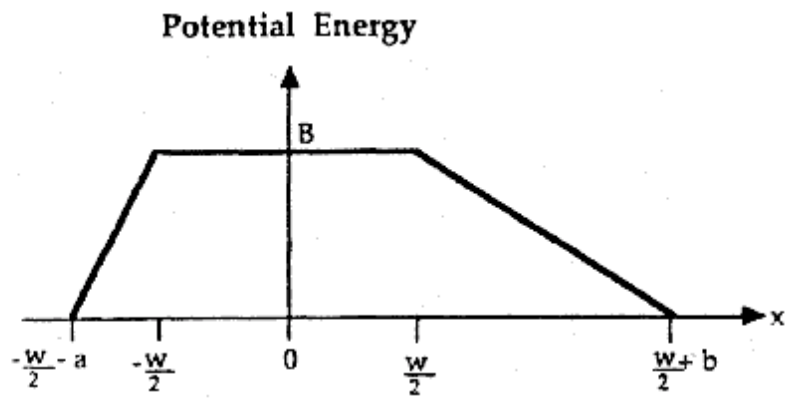

Potential Energy

Fig. 1. Trapezoidal potential energy barrier of height *B*.

For a one-dimensional square barrier of width *w* and potential energy (height) *B* (Figure 1 with vertical sides, i.e., $a = b = 0$) solution of the Schrödinger equation yields the transmission coefficient:

$$TT^* = \left[1 + \frac{B^2 \sinh^2 \beta w}{4E(B-E)}\right]^{-1} \qquad (2)$$

where $\beta = [2m(B-E)]^{1/2}/\hbar$, E is the energy of the incident particle, and *m* is its mass. Note that quantum mechanically where $E < B$, the momentum of the particle is imaginary and its kinetic energy is negative inside the barrier. As $w \to 0$, $TT^* \to 1$ for finite *B*, and as $w \to \infty$, $TT^* \to 0$. Similarly, as $E \to 0$, $TT^* \to 0$, for $w \neq 0$. When w = 0, $TT^* = 1$, as $E \to 0$. For any general barrier, the transmission coefficient *TT\** and amplitude *T* are the same in all direction for a free particle (cf. Appendix B). The reflection coefficient has the same magnitude for all directions of incidence on the barrier, but the reflection amplitudes do not all have the same phase (cf. Appendix B).

### 3. CLASSICAL HIGH JUMPING

Now let us compare the above quantum tunneling results with classical high jumping for a nonabsorbing trapezoidal barrier of height (maximum potential energy) *B*, top width *w*, and total width *w + a + b* as defined by Figure 1. This is a fairly general barrier, as, by appropriate choice of *w*, *a*, and *b*, a square, triangular, or line barrier can also be represented. Solutions may be readily obtained by conservation of energy, as shown here, though any of a number of methods may be used.

Equating the energy E of the rope in the zero potential energy region to its potential energy when it can just barely get over the barrier, we have in general $E = \int_0^L \rho V dx$, where $\rho(x)$ is the appropriate linear density (mass density for a gravitational field, charge density for an electric field, etc.) of the rope of length L for the particular field. For the potential V - V(x) of Figure 1 ($V_o$ is the maximum potential) there are three solution regimes: $E_{Short}$ for $0 < L < w$; $E_{Medium}$ for $w < L < w + a + b$; and $E_{Long}$ for $L > w + a + b$:

$$E_S = \int_0^L \rho V_o dx \tag{3}$$

$$E_M = -\int_0^{L_1} \rho x(V_o/a)dx + \int_0^L \rho V_o dx - \int_0^{L_1} \rho x(V_o/b)dx \tag{4}$$

$$E_L = -\int_0^{L_1} \rho x(V_o/a)dx + \int_{G_1+G_2}^L \rho V_o dx - \int_0^b \rho x(V_o/b)dx \tag{5}$$

$L_1$ and $L_2$ are the equilibrium lengths of the medium rope which extend over either side of the top of the barrier. In the long-rope case, $G_1$ and $G_2$ are the lengths of the rope on either side of the barrier edges at zero potential.

The solutions to equations (3) - (5) for the energy that the body must have to just clear the barrier of Figure 1 are

$$E_S = B, \qquad\qquad 0 \leq L < w \tag{6}$$

$$E_M = B - B(L-w)^2/2(a+b)L, \qquad w \leq L \leq w+a+b \tag{7}$$

$$E_L = B[2(a+b)/2L + w/L], \qquad w+a+b \leq L \tag{8}$$

The solutions (6) - (8) give the minimal energy E that a rope of length L needs to clear (penetrate) the barrier. Just as in the dase of quantum tunneling, it is possible to clear the barrier even if E < B. Since the solutions are symmetric in a and b, high jumping is the same in both directions, as is tunneling (cf. Appendix B). Another of many similarities is that when $w \to \infty$, E must be $\geq B$ to go over the barrier. Of course, when $0 \leq L \leq w \neq 0$, $E \geq B$ for clearing the barrier.

As can be seen from Appendix A, there are many more interesting correlations with quantum mechanics at even this stage of the analysis. So far, we have established that it is possible for a body whose energy is less than its potential energy at the top of a barrier to cross (high jump) the barrier, provided the body is an extended object, as an analog to quantum mechanics. However, the high-jumping coefficient at this stage is only 0 or 1. As can be seen in the next section, the analog may be taken even further.

## 4. LINK BETWEEN CLASSICAL AND QUANTUM BARRIER PENETRATION

The probabilistic nature of quantum tunneling is recovered if we assume that there is a distribution of rope lengths. Equations (7) and (8) should be interpreted as relationships giving *L*, the minimum rope length needed for barrier penetration, in terms of *E*, *B*, *w*, and *a + b*. For simplicity, consider the square barrier with $a = b = 0$; then for $E < B$ equation (8) is operative and can be rewritten as

$$L = Bw / E \qquad (9)$$

Only those ropes with length $x \geq L$ as given by equation (9) will go over the barrier, and those smaller than *L* will not. One would anticipate that in order to match the quantum solution, the de Broglie wavelength $\lambda = h / (2mE)^{1/2}$, is the basic length-scale parameter for such distributions. Remarkably, we do find that the distributions *f(x)* of rope lengths are distributed about $\lambda$.

A normalized distribution we find, with a most probable value at $\lambda \sqrt{4/3}$, as shown in Figure 2:

$$f(x/\lambda) = (x/2\lambda)\left[1 + \left(x^2 / 4\lambda^2\right)\right]^2 \qquad (10)$$

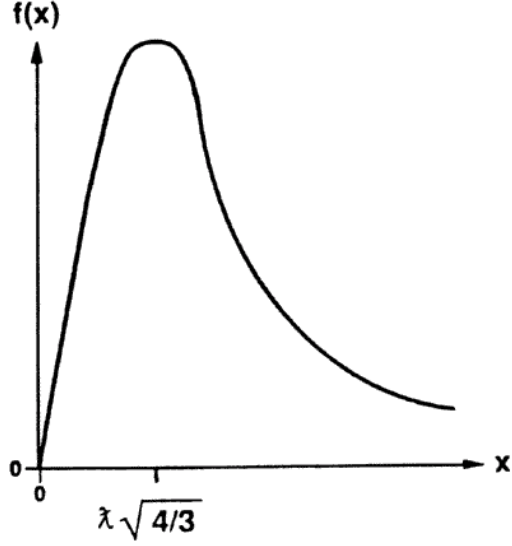

Fig. 2. Distribution of rope lengths.

The classical transmission coefficient $T_c$ for this distribution is

$$T_c = \int_L^\infty f(x)dx \Big/ \int_0^\infty f(x)dx = \int_{L/\lambdabar}^\infty f(x/\lambdabar)d(x/\lambdabar) \qquad (11)$$

The solution to equation (11) is

$$T_c = \left[1 = \left(L^2/4\lambdabar^2\right)\right]^{-1} = \left[1+(B\beta w)^2/4E(B-E)\right]^{-1} \qquad (12)$$

Equation (12) agrees very well with the quantum mechanical $TT^*$ of equation (2) in the domain where , which one would expect by the correspondence principle. For any $B$ there is close agreement between $T_c$ and $TT^*$, where $(2mBw^2)^{1/2} < \hbar$, and there is exact agreement in the limit as the positive quantity $(B-E)$ goes to zero:

$$T_c = TT^* = \left[1+\left(mBw^2/2\hbar^2\right)\right]^{-1} \qquad (13)$$

## 5. CONCLUSION

Thus, if the assumption "that the body length is zero" is not made for a body incident on a barrier, then "tunneling" need no longer be considered a purely quantum effect, as classical "high jumping" gives analogous results. Based upon the analysis in this paper, it can be seen directly how it is possible for a body to appear on the other side of a barrier even when its incident energy is less than its potential energy at the top of the barrier. Although the entire body goes over the barrier, its center of mass and/or center of interaction literally go through the barrier.

This model illustrates that a classical system can exhibit what were thought to be quintessential quantum mechanical properties when its energy is spatially distributed over a sufficiently large distance and is *coherently* coupled. In the macroscopic world, the body length may take on a large range of values as an independent variable. In the microcosm, only those body lengths distributed around the de Broglie wavelength lead to reasonable agreement with the quantum transmission coefficient.

**APPENDIX A. INTERESTING FEATURES OF ROPE SOLUTIONS**

Even before a statistical aspect is introduced, there are a number of interesting features of the simple rope solutions that may be nonintuitive.

1. For L > 0, as the barrier gets infinitesimally thin, i.e., ; it is possible to penetrate the barrier even as $E \to 0$. At first sight this classical result may be surprising and even appear a bit quantum mechanical. It may be understood by realizing that as the barrier gets thinner for a fixed length of rope *L*, the fraction of the rope's kinetic energy that is converted to potential energy gets correspondingly smaller. In the limit of an infinitesimally thin barrier ( a line barrier), this fraction of the rope's energy goes to zero. Thus the rope can make it over a large but finite barrier of height *B*, even though the rope has an arbitrarily small amount of energy. This is the same as the quantum limit for an infinitesimally thin barrier of finite height as $E \to 0$, since $TT^* \to 1$.

2. As in quantum mechanics, for a finite barrier of width w > 0, when E = 0, classical high jumping cannot clear the barrier. However, *E* can get indefinitely small (but not =0) in successfully clearing the barrier as $L \to \infty$ (holding the total charge, mass, etc., fixed as would be the case for a fundamental particle). The reason is similar to that of feature 1, because the fraction of the rope's energy that is potential energy $\to 0$ as $L \to \infty$. For quantum tunneling, whenever w > 0, $TT^* \to 0$ as $E \to 0$.

3. For a point body of L = 0, $E \geq B$ to clear the barrier. However, the limit of $L \to 0$ approaching a point body has unexpected solutions. With $w = 0$ as , equation (7) reduces to B(1 - f/2) in the limit as $L \to 0$, where f = L/(a + b) and $0 \leq f \leq 1$. This yields . For $f \geq 1$, equation (8) reduces to , yielding . With a + b = 0 as $w \to 0$, equation (6) or (7) gives E = B for L < w. However, for , equation (8) reduces to E = gB and thus 0 < E < B.

## APPENDIX B. SYMMETRY AND ASYMMETRY OF QUANTUM TUNNELING

Establishing some general properties of quantum mechanical tunneling facilitates a comparison with the classical high-jumping results. Consider a *general* one-dimensional barrier of potential energy $V(x)$ between two regions of zero potential. We have the following solutions of the Schrödinger equation (S-eq.) on the incident (Region 1) and transmitted sides (Region 2) of the barrier.

Region 1:

$$\Psi(x) = e^{ikx} + Re^{-ikx} \qquad (B1)$$

Region 2:

$$\Psi(x) = Te^{ikx} \qquad (B2)$$

where *k* is the wave vector, $RR^*$ is the reflection coefficient, and $TT^*$ is the transmission coefficient. For a wave incident on the barrier in the opposite direction $\psi(x) = \Psi(-x)$ is the solution of the S-eq. with V(-x) replacing V(x). The solutions in regions 1 and 2 are as follows:

Region 1:

$$\psi(x) = Te^{ikx} = \Psi(-x) \tag{B3}$$

Region 2:

$$\psi(x) = e^{ikx} + Re^{ikx} = \Psi(-x) \tag{B4}$$

For a nonabsorbing barrier, *V* is real. Therefore a solution of the S-eq. with *V(x)* is

$$\Psi(x) = [-R*/T*]\psi(x) + [1/T*]\psi*(x) \tag{B5}$$

where, for Region 1

$$\Psi(x) = e^{ikx} + [-R*T/T*]e^{ikx} \tag{B6}$$

and for Region 2

$$\Psi(x) = [(1 - RR*)/T*]e^{ikx} = Te^{ikx} \tag{B7}$$

Therefore the transmission amplitude *T* and the phase are the same in the two directions, i.e., quantum mechanical tunneling is symmetric with respect to direction. Although the reflection amplitude *R*, in Region 2 has the same magnitude *R* as in Region 1, $R_2 = -R*T/T* \to |R|$, it does not in general have the same phase.

**APPENDIX C. DISCUSSION OF CLASSICAL PENETRATION**

Our concept of particles as extended objects having lengths distributed about the de Broglie wavelength warrants consideration with respect to descriptions of particles as "waves and particles simultaneously," compared with complementarity. The direct statistical nature of barrier penetration in our model warrants further study and reflection with respect to the possibility of directly detecting de Broglie waves (Garuccio *et al.*, 1981). In fact, the length scale corresponding to the de Broglie length ($\lambda$ here) may be inferred from the distribution of rope lengths as related to the high-jumping energy. This is analogous to the relationship between the Bohr radius and the energy levels of the Bohr model of the atom, where the Bohr radius is equal to $\lambda$ for the ground state of a hydrogenic atom.

At the very least, classical high jumping can provide a physical insight into quantum tunneling. This may have value in the sense that if Bohr had created the Bohr model of the atom after atomic energy levels had already been calculated from the Schrödinger equation, his model would still provide valuable insights into what would otherwise be an abstract mathematical process.

One difficulty of the classical rope analysis is that it is discontinuous at $E = B$, requiring all particles to penetrate $E > B$. However, the quantum result is continuous at $E = B$, with partial reflection for $E > B$. The formal extension of equation (12) into the Region $E > B$ does approximate the general trend of the quantum solution, though it does not contain its oscillatory character. A more complete analysis should consider the combined effects of varying rope lengths and internal energy conversion. The latter could result when the leading edge of the rope is incident on the barrier and kinetic energy of translation would couple to internal energy, and thus account for reflection when $E \geq B$. In the one-dimensional case, the internal energy can only be oscillational. In two or more dimensions, the internal energy could also be rotational.

When quantum mechanics deals with an extended body with distributed mass, the additional terms in the Hamiltonian related to the potentially new degrees of freedom such as oscillation and rotation about the center of mass are represented in the Schrödinger equation. Even if they were to be included, it is not clear that their contribution would materially affect the solution. For example, for objects such as an electron, rotational excited states require a high energy input for excitation. Hence they are not present under ordinary circumstances. A barrier only appears immutable when a model is constructed of a point particle interacting with a given barrier.